\documentclass[12 pt, article]{article}
\usepackage[left=1in, right=1in, top=0.75in, bottom=0.75in]{geometry}
\usepackage[utf8]{inputenc}
\usepackage{setspace}
\usepackage{graphicx}
\usepackage{authblk}
\usepackage{sectsty}
\usepackage{gensymb}
\usepackage{hyperref}
\usepackage{parskip}
\usepackage{caption}
\usepackage{titlesec}
\usepackage{wrapfig}
\usepackage{xcolor}
\usepackage[utf8]{inputenc}
\graphicspath{{images/}}
\sectionfont{\fontsize{13}{16}\selectfont}

\titlespacing\section{0pt}{12pt plus 2pt minus 2pt}{12pt plus 2pt minus 2pt}
\usepackage[backend=biber, style=nature]{biblatex}
\newcommand{\ringonemacro}{$I_{0\overline{1}10} + I_{\overline{1}010} + I_{\overline{1}100}\,$ }
\newcommand{\ringtwomacro}{$I_{\overline{2}110} + I_{1\overline{2}10} + I_{11\overline{2}0}\,$ }
\title{\Large\textbf {Local atomic stacking and symmetry in twisted graphene trilayers}}

\author[1]{Isaac M. Craig}
\author[1]{Madeline Van Winkle}
\author[1]{Catherine Groschner}
\author[1]{Kaidi Zhang}
\author[1]{Nikita Dowlatshahi}
\author[2]{Ziyan Zhu}
\author[3]{Takashi Taniguchi}
\author[4]{Kenji Watanabe}
\author[5,6]{Sinéad M. Griffin}
\author[1,7,*]{D. Kwabena Bediako}
\affil[1]{\textit{Department of Chemistry, University of California, Berkeley, CA 94720, USA}}
\affil[2]{\textit{SLAC National Accelerator Laboratory, Stanford, CA, USA}}
\affil[3]{\textit{International Center for Materials Nanoarchitectonics, National Institute for Materials Science, 1-1 Namiki, Tsukuba 305-0044, Japan}}
\affil[4]{\textit{Research for Functional Materials, National Institute for Materials Science, 1-1 Namiki, Tsukuba 305-0044, Japan}}
\affil[5]{\textit{Molecular Foundry, Lawrence Berkeley National Laboratory, Berkeley, CA 94720, USA}}
\affil[6]{\textit{Materials Sciences Division, Lawrence Berkeley National Laboratory, Berkeley, CA 94720, USA}}
\affil[7]{\textit{Chemical Sciences Division, Lawrence Berkeley National Laboratory, Berkeley, CA 94720, USA}}
\affil[*]{Correspondence to: bediako@berkeley.edu}

\date{}
\begin{document}
\maketitle

\doublespacing
\textbf{Abstract}

Moiré superlattices formed by twisting trilayers of graphene are a useful model for studying correlated electron behavior and offer several advantages over their formative bilayer analogues, including a more diverse collection of correlated phases and more robust superconductivity. Spontaneous structural relaxation significantly alters the behavior of moiré superlattices, and has been suggested to play an important role in the relative stability of superconductivity in trilayers. Here, we use an interferometric four-dimensional scanning transmission electron microscopy approach to directly probe the local graphene layer alignment over a wide range of trilayer graphene structures. Our results inform a thorough understanding of how reconstruction modulates the local lattice symmetries crucial for establishing correlated phases in twisted graphene trilayers, evincing a relaxed structure that is markedly different from that proposed previously. 

\newpage

Since their initial discovery, graphene-based moiré superlattices have emerged as valuable tools for investigating the balance between key parameters in strongly correlated phases \cite{balents2020superconductivity, cao2018correlated,cao2018unconventional,lau2022reproducibility,tian2023evidence}. Their effectiveness stems, in part, from the diverse phenomena they manifest when adjusting twist angle, carrier density, and thickness. Notably, trilayer graphene showcases markedly distinct properties across its native stacking arrangements \cite{bao2011stacking, kumar2011integer, lui2011observation, yuan2011landau}. Bernal-stacked (ABA) trilayer graphene is a semi-metal and features poorly coupled bands resembling a composite of monolayer and bilayer graphene \cite{craciun2009trilayer}. Conversely, rhombohedral-stacked (ABC) trilayer graphene displays hybridization among all three layers, Mott insulating states \cite{chen2019evidence}, metallic behavior \cite{zhou2021half}, and superconductivity \cite{zhou2021superconductivity}. The variations in local lattice symmetry significantly contribute to the realization of distinct properties in few-layer graphene systems, both within high-symmetry structures and within the locally ordered domains of moiré superlattices more broadly \cite{koshino2009gate, morell2013electronic}.

Particularly noteworthy is the characterization of flat bands in twisted bilayer graphene (TBLG) as a fragile topological phase protected by space-time inversion symmetry \cite{ahn2019failure, song2019all}. This protection persists despite the atomic lattice's inversion symmetry being limited to instances where the carbon atoms in each layer align vertically (AA stacking)  \cite{zou2018band}. The extent of overlap between localized states within the AA regions that are associated with the flat bands, and consequently the size of AA stacking regions, is believed to significantly influence the superconducting current and transition temperature in twisted structures \cite{torma2021superfluidity}. Twisted graphene multi-layers, including magic angle twisted trilayer graphene (MATTLG), rely on the same $C_{2z}T$ symmetry as their bilayer counterparts \cite{mora2019flatbands}. Similar to twisted bilayers, twisted trilayer graphene (TTLG) structures exhibit local $C_{2z}T$ symmetry in pockets of AAA-type alignment. Recent findings indicating that superconducting phases in MATTLG are more resilient to variations in twist angle and gating than those observed in bilayers \cite{park2021tunable, hao2021electric, kim2022evidence, zhang2021correlated} have prompted intrigue regarding the potentially contrasting impact spontaneous lattice relaxation may play in these two systems. While TBLG spontaneously undergoes a reduction in the proportion of AA stacking \cite{yoo2019atomic,kazmierczak2021strain}, it has been suggested that relaxation in trilayers may instead augment the prevalence of inversion symmetric alignments \cite{turkel2022orderly}. This underscores the importance of precise structural characterization to uncover the intricacies of spontaneous lattice relaxation in twisted graphene trilayers.

Here, we employ an interferometric methodology based on four-dimensional scanning transmission electron microscopy (4D-STEM \cite{ophus2019four}) known as Bragg interferometry (Fig. 1A) \cite{kazmierczak2021strain, van2022quantitative, zachman2021interferometric}. This technique leverages the local interference pattern in diffracted electron beams to unambiguously deduce the stacking orientation of atomic layers. Unlike scanning tunneling microscopy (STM) and more conventional  STEM methods, this technique allows us to probe moiré patterns within encapsulated materials. Importantly, it also facilitates the selective imaging of individual bilayer interfaces within complex multilayered materials. These results offer a direct assessment of the local atomic stacking within twisted graphene trilayers.  We find that the results of this 4D-STEM measurement suggest a picture of reconstruction that is markedly different from that proposed by previous STM work \cite{turkel2022orderly}, and one that is consequential for understanding the correlated electron physics in these materials.

The interferometric 4D-STEM technique we use involves scanning a converged electron beam across the sample of interest and collecting individual diffraction patterns for each real space position of the probe (Fig. 1A). Throughout this work, we use the notation shown in Fig. 1B to label the twist angles, $\theta$, within the trilayer sample. Here, $\theta_{12}$ and $\theta_{23}$ denote the twist angles between layers 1 \& 2, and layers 2 \& 3 respectively such that $\theta_{13}$ = $\theta_{12} + \theta_{23}$. We further use the labels shown in Fig. 1C to denote the various high-symmetry stacking configurations realized within the moiré. The converged beam electron diffraction (CBED) patterns collected at each probe location then appear as shown in Figs. 1D,E, where each layer of the material generates a set of Bragg disks. The overlap between Bragg disks originating from different layers (inset Figs. 1D,E) is then used to determine the stacking orientation of those two layers. As an example, Figs. 1F,H shows how the intensity of the overlap between layers 1 and 3—denoted as `(1,3) overlap'—varies across the sample. This modulation in intensity directly manifests the moiré pattern between layers 1 and 3 but is insensitive to the orientation of the second layer. Similarly, the variation in the intensity of the (1,2,3) overlap region (Figs. 1G,I) reveals the modulation in stacking order between all three graphene layers. 

Therefore, by exploiting the relationship between stacking order and overlap region intensity (see Supplemental Information sections 4--6), we map the variation in atomic stacking and hence reconstruction within trilayer graphene samples. Results of these analyses are shown in Fig. 2 for a structure that we call `AtA' in which the top and bottom graphene layers are perfectly aligned to each other and twisted with respect to the middle layer (Fig. 2B). In this structure, the average intensity of  the overlap regions in the first ring of Bragg reflections and the average overlap region intensity in the second ring of reflections can be used to determine the local stacking configuration. Using the bi-variate color-legend shown in Fig. 2A (in which the high symmetry stacking configurations associated with each color are overlaid), we create a map of local atomic stacking within an AtA sample (Fig. 2C). 

The local atomic stacking shown in Fig. 2C indicates that this particular AtA sample for which ($\theta_{12}=1.05\degree$) relaxes to decrease the total amount of AAA stacking (white) when compared to the stacking order distribution in a rigid AtA trilayer (Fig. 2E). This is expected as AAA stacking is roughly 29.5 and 36.5 meV/unit cell higher in energy than A-SP-A and ABA stacking respectively (See Supplemental Information section 9). Further, the histogram shown (Fig. 2D) illustrates that this sample contains considerably more ABA, BAB, and SP type stacking than AAA type stacking (See Supplemental figure 3 for the stacking histogram expected of a rigid sample). This reconstruction also manifests in the  mean line-cut  shown in Fig. 2F, which corresponds to the  average over all line-cuts equivalent to that denoted by the  dotted line in Fig. 2C. From this profile, it is evident that the widths of the  AAA regions are smaller than expected for a rigid trilayer, which is robust the presented standard deviation (shaded region) and noise-driven differences in normalization (See Supplemental Information section 7) as well as and limitations in spatial resolution due to appreciable beam-width biasing (See Supplemental Information section 6). 

These results, as well as results for other twist angles shown in Supplemental Information section 8, together illustrate that AtA trilayers show an observably reconstructed atomic stacking distribution up to a at least a twist angle of $1.81\degree$, with few differences seen between the $1.81\degree$ and $1.0\degree$ samples. We also observe that the AtA trilayers show a pattern of reconstruction similar to that of a twisted bilayer within this twist angle range \cite{yoo2019atomic,kazmierczak2021strain}, although a more quantitative comparison between the bilayer and AtA trilayer reconstruction necessitates more detailed intensity fitting approach to map strain tensor fields \cite{kazmierczak2021strain,van2022quantitative} and will be addressed in future work. 

A similar analysis is carried out for samples which we refer to as `tAB', in which the bottom two graphene layers are aligned AB and the top layer is twisted (Fig. 2H), creating a structure sometimes referred to as a `monolayer-on-bilayer graphene', which also exhibits correlated electron physics \cite{tong2022spectroscopic,park2020gate}. Unlike the AtA trilayers, these tAB structures show atomic reconstruction patterns driven by a preference to decrease the relative portion of AAB stacking,  as seen by comparing Fig. 2I to the rigid stacking order distribution (Fig. 2K), the histogram in Fig. 2J, and the corresponding line-cut in Fig. 2L (see Supplemental Information section 7 for normalization bias and error margins),  which illustrates the tAB structure reconstructs such that the portions of AAB and BAB stacking are no longer equivalent as they would be in a rigid moiré. This decrease in AAB stacking is expected, it is 17.9 meV/unit cell higher in energy than ABC and BAB (See Supplemental Information section 9). We note here that this manifests in the  AAB  regions appearing to have a lower peak \ringtwomacro intensity to the  BAB  regions, while these regions are expected to appear sharper but with similar maximum intensity. This is likely due to broadening from a number of factors associated with measurement acquisition and post-processing, especially the beam-width biasing and the Gaussian filter used (see Supplemental Information section 6). 

The observation that  AAB and BAB  regions are observably distinct even at a twist angle of $1.4\degree$ is nonetheless notable. This effect is more dramatic at smaller twist angles as seen in the stacking order percent area trends and maps gathered within the $0.1\degree$ -- $1.5\degree$ twist angle regime. While the approach used in this work prohibits a quantitative comparison of the  AAB and BAB  domain sizes (to be addressed in future work), these results still clearly establish that the size of ABC and  BAB  domains are comparable to each-other and much larger than the  AAB  domains. Moreover, the stacking order distributions seen for tAB appear similar to those observed in twisted bilayer graphene \cite{kazmierczak2021strain}, suggesting that atomic reconstruction in tAB trilayer samples can be explained primarily from consideration of the twisted interface. 

The results discussed thus far concern a limiting case of graphene trilayers wherein two of the layers are perfectly aligned. While these materials are conceptually simpler and display rich physical properties which merit their investigation, this interferometric 4D-STEM technique also permits us to study a broader array of twisted trilayer structures with two independent twist angles. In these more complex multilayered samples, the ability to selectively probe buried bilayer interfaces allows us to independently image the larger scale moiré pattern and evaluate its effect on local stacking order. 

Following this approach, we extract stacking order maps associated with the larger moiré pattern from double overlap (Fig. 3A) and triple overlap (Fig. 3B,E). These are compared to the maps calculated for rigid moirés in Fig. 3C. For the the double overlap case, Fig. 3A reveals the presence of large local regions in which two layers are aligned directly atop each-other (AtA or tAA, white) and regions in which two layers are aligned AB (AtB or tAB, blue). From comparing the stacking distributions of samples with $0 < \theta_{13} \ll \theta_{23}$ (three leftmost panels in Fig. 3A, illustrated in Fig. 3D) and $0 < \theta_{23} \ll \theta_{13}$ (rightmost panel in Fig. 3A, illustrated in Fig. 3F), we find that the two regimes display distinct reconstruction patterns. When $\theta_{13} \ll \theta_{23}$, the observed atomic reconstruction is driven by a slight preference for AtA type stacking (white) over AtB (blue) and soliton-type (grey) regions. This result is somewhat unexpected as the energetic difference between rigid AtA and AtB domains (driven only by inter-layer coupling between the top and bottom layers) has been previously presumed to play a minor role in reconstruction \cite{zhu2020twisted}. Moreover, previous STM studies \cite{turkel2022orderly} concluded that trilayers with $0 < \theta_{13} \ll \theta_{23}$ relaxed to effectively eliminate AtB domains.

Fig. 3A also shows (rightmost panel) that the atomic reconstruction pattern for $\theta_{23} \ll \theta_{13}$ is instead driven by a preference to minimize the high energy tAA (white) domains, within which every possible stacking configuration must place two carbon atoms from neighboring layers directly atop each other — an arrangement that is sterically unfavorable \cite{yoo2019atomic,kazmierczak2021strain}. The extent of reconstruction in these $\theta_{23} \ll \theta_{13}$ samples is therefore much larger, since the energy difference between rigid tAA vs tAB domains ($\approx 18.2$ meV/unit when considering only adjacent interfaces) is much larger than that between rigid AtA vs AtB domains ($\approx 0$ meV/unit when considering only adjacent interfaces) \cite{zhu2020twisted, kim2022evidence}. This is reflected in the difference between the structures shown in the second and fourth columns of Fig. 3, in which both structures have comparable twist angles, but the structure in the fourth panel is observably more reconstructed, with the tAA domains appearing as a highly contracted spot. This spot appears orange rather than white due to beam-width and data processing effects (see Supplemental Information section 6). Although the weaker higher frequency texture observed within the white and blue domains in Fig. 3A might arise from the smaller scale moiré pattern imparting a modulation in these stacking distributions, this pattern likely predominantly results from a small bleed-in of the (1,2,3) interference pattern, which is hard to completely exclude with virtual apertures while retaining sufficient signal-to-noise ratios.

After extracting the local AtA/tAA and AtB/tAB domains as shown in Fig. 3A, we now examine the (1,2,3) overlap region, which is associated with all three graphene layers (Figs. 3B), to understand how the smaller scale moiré pattern modulates local stacking order within these larger domains (representative regions of these maps are magnified in Fig. 3E). Additional maps are shown in Supplemental Figs. 7-8. We note that, as noted in previous work \cite{turkel2022orderly}, we see only two clear periodicities in our data despite the presence of three moiré wavelengths from each twisted bilayer interface. However, this does not necessarily imply that only two moiré wavelengths are present; inspection of the atomic stacking maps expected from even rigid structures (Fig. 3C and Supplemental Figures 7-8) reveals that the smaller and larger periodicities observed reflect only the largest and smallest twist angles, respectively.

Taken together, the data in Fig. 3 allow a quantification of the area fractions in TTLG samples and the development of a qualitative model for reconstruction in the limit of slight misalignment (Fig. 4). Fig. 4A shows that for the larger moiré pattern, the proportion of AtA/tAA and AtB/tAB stacking domains inverts across the regimes illustrated in Figs. 3D,F. Associated area fractions from our measurements and those from continuum relaxation simulations as a function of $\theta_{13}-\theta_{23}$ are shown in Fig. 4A. Experiment and simulation show good agreement in the overall trends, though the measurements show a more gradual decline in area fraction of AtA/tAA (and corresponding rise in that of AtB/tAB/SP) than the simulation with increasing $\theta_{13}-\theta_{23}$. This slight discrepancy may arise because of kinetic effects preventing the system from realizing the theoretically optimal extent of relaxation driven by layers not immediately adjacent. 

For the smaller scale moiré superlattice, we find that this pattern appears relatively invariant within the AtA, AtSP, AtB, and tAB domains (SI Section 12). Indeed, the measured proportion of stacking orders within the AtA regions of Fig. 3  is very similar to the pure AtA sample seen in Fig. 2C, suggesting that the larger scale moiré plays a relatively minor role in the reconstruction of the smaller scale moiré. We do however observe some differences between the smaller-scale moiré within different domains. As shown in Fig. 4B, measurements of local $\theta_{12}$ values within the AtA and AtB domains of $\theta_{13} \ll \theta_{23}$ samples display a slightly smaller $\theta_{12}$ angle within AtA regions as compared to the values in adjacent AtB domains. This tightening of the smaller-scale moiré within AtB regions might help facilitate the overall minimization of these AtB regions. 

Lastly, we investigate the maps of local atomic stacking order in regions with an increasingly large extent of extrinsic heterostrain, $\epsilon$. From the maps shown in Fig. 5, we find that extrinsic heterostrain acts predominantly on the larger scale moiré pattern and has a diminishing effect on the smaller scale moiré superlattice, consistent with previous work on bilayer moiré systems \cite{kazmierczak2021strain,van2022quantitative}.
Notably, in the most heterostrained sample of Fig. 5, despite similar $\theta_{13}$, the islands of AtA are deformed into stripes. These features have also been previously visualized in STM studies and attributed to heterostrain between the top and bottom layers \cite{kim2022evidence}. Heterostrain is therefore a powerful tuning knob for manipulating the contiguity of AtA domains (from islands to stripes) at the expense of AtB regions, potentially modulating the emergence of correlated phases that rely on the $C_{2z}T$-symmetric AtA domains. 

In conclusion, the nature of atomic reconstruction unveiled here for twisted trilayers is markedly different than that proposed in previous work, wherein it was suggested that slightly misaligned MATTLG samples relax to almost exclusively AtA regions, with the AtB and SP regions stretched into thin domain boundaries and/or topological defects that contribute insignificantly to the STM measurements \cite{turkel2022orderly}. While Fig. 2 shows that at length scales where only one moiré wavelength is apparent (when $\theta_{13} \approx 0\degree$), trilayers do favor the formation of large AtA domains, and in that case the local structure of trilayers is driven primarily by consideration of the smaller moiré, we see a clear presence of considerable AtB type stacking down to $\theta_{13} = 0.20\degree$ (Fig. 3). This observation contrasts previous claims of trilayer samples containing contributions from only AtA regions at a $\theta_{13}$ of $\approx 0.25\degree$,  with further discussion of our results in the context of these prior measurements provided in Supplemental Information section 10. Taken together, our measurements highlight the particular utility of interferometric 4D-STEM imaging alongside other scanning probe techniques like STM for characterizing complex multi-layered moirés, as the ability to apply a direct structural probe selectively to separate interfaces can uncover the complex picture of atomic reconstruction. 

The extent of AtB stacking observed could have major implications for understanding superconductivity in misaligned MATTLG \cite{hao2021electric,kim2022evidence} and recently discovered moiré quasicrystal systems \cite{uri2023quasicrystal}. For instance, if $\theta_{13} \ll \theta_{23}$ configurations such as MATTLG favored entirely AtA stacking as previously proposed, their correlated behaviors could be predominantly understood by consideration of the $C_{2z}T$ inversion symmetric AAA stacking regions much like TBLG. While our measurements support a relative contraction of AtB domains, it is no where near as dramatic, revealing that sizable AtB portions remain following reconstruction.  These significant AtB domains may instead suggest that the ABC, AAB, and ABB regions, which have been shown to host correlated electronic phases \cite{zhou2021superconductivity,chen2020tunable,xu2021tunable,li2022imaging} despite a lack of inversion symmetry, may play an important role in understanding correlated electron physics in some twisted trilayers. 

\section*{Acknowledgements}
We thank P. Kim, J. Ciston, K. Bustillo, and C. Ophus for vis-a-vis and epistolary discussions. This material is based upon work supported by the US National Science Foundation Early Career Development Program (CAREER), under award no. 2238196 (D.K.B). I.M.C. acknowledges a pre-doctoral fellowship award under contract FA9550-21-F-0003 through the National Defense Science and Engineering Graduate (NDSEG) Fellowship Program, sponsored by the Air Force Research Laboratory (AFRL), the Office of Naval Research (ONR) and the Army Research Office (ARO). C.G. was supported by a grant from the W.M. Keck Foundation (Award no. 993922). Experimental work at the Molecular Foundry, LBNL was supported by the Office of Science, Office of Basic Energy Sciences, the U.S. Department of Energy under Contract no. DE-AC02-05CH11231. Confocal Raman spectroscopy was supported by a Defense University Research Instrumentation Program grant through the Office of Naval Research under award no. N00014-20-1-2599 (D.K.B.). Other instrumentation used in this work was supported by grants from the Canadian Institute for Advanced Research (CIFAR–Azrieli Global Scholar, Award no. GS21-011, D.K.B), the Gordon and Betty Moore Foundation EPiQS Initiative (Award no. 10637, D.K.B), and the 3M Foundation through the 3M Non-Tenured Faculty Award (no. 67507585, D.K.B). Theoretical work was supported by the Theory of Materials FWP at Lawrence Berkeley National Laboratory, funded by the U.S. Department of Energy, Office of Science, Basic Energy Sciences, Materials Sciences and Engineering Division, under Contract No. DE-AC02- 05CH11231 (S.G). K.W. and T.T. acknowledge support from JSPS KAKENHI (Grant Numbers 19H05790, 20H00354 and 21H05233).

\section*{Author Contributions}
I.M.C., M.V., C.G., and D.K.B. conceived the study. M.V., C.G., K.Z., and N.D. fabricated the samples. M.V., C.G., and I.M.C. performed the experiments.
I.M.C. and Z.Z. performed the continuum simulations using code developed by Z.Z. I.M.C. developed and implemented the interferometry code (with assistance from C.G.) and analyzed the data. T.T. and K.W. provided the hBN crystals. D.K.B. and S.M.G. supervised the work. I.M.C. and D.K.B. wrote the manuscript with input from all co-authors. 

\section*{Competing Interests}
The authors declare no competing interests.

\section*{Methods}
\subsection*{Sample Preparation} All graphene trilayers were fabricated using the tear-and-stack technique \cite{cao2018correlated,cao2018unconventional}. Briefly, a polybisphenol-A-carbonate/polydimethylsiloxane (PC/PDMS) stamp was first used to pick up 5–10 nm hexagonal boron nitride (hBN) off a \(\mathrm{SiO_2}\)/Si substrate. This hBN was then used to pick up and tear graphene monolayers and/or bilayers, also on \(\mathrm{SiO_2}\)/Si. The remaining graphene portions were then sequentially rotated and picked up to construct the desired trilayer structure, which was stamped onto a Norcada TEM grid (200 nm silicon nitride with 2 \(\mathrm{\mu m}\) holes).

\subsection*{Electron Microscopy Measurements} All electron microscopy measurements were performed at the National Center for Electron Microscopy at Lawrence Berkeley National Laboratory. Low-magnification dark-field TEM images were collected using a Gatan UltraScan camera on a Thermo Fisher Scientific Titan-class microscope at 60 kV to identify regions of interest prior to 4D-STEM acquisition. 4D-STEM data were obtained using a Gatan K3 direct detection camera and Gatan Continuum imaging filter on the TEAM I microscope (aberration-corrected Thermo Fisher Scientific Titan 80–300). We operated in energy-filtered STEM mode at 80 kV using an 10-eV energy filter centered around the zero-loss peak, convergence angle of 1.71 mrad, and a typical beam current of 45–65 pA for an overall effective probe size of 1.25 nm corresponding to the full-width at half-maximum value. The diffraction patterns collected correspond to a step size of 0.5-2 nm depending on the sample. We operated the K3 camera in full-frame electron counting mode with a binning of 4 × 4 pixels, energy-filtered STEM camera length of 800 mm, and exposure time of 13 ms which was the sum of multiple counted frames. Considerations for choice in convergence angle and camera length are discussed in Supplemental Information section 2.  

\subsection*{Virtual Aperture Selection and Stacking Order Maps}
To obtain the virtual dark fields shown in Figure 1 of the main text, we first had to isolate the regions of reciprocal space associated with diffraction of (1,2,3) and (1,3) layers. First the individual Bragg disks are attributed to layers as described in Supplemental Information section 3, using the known order in which each graphene layer was picked up, optical micrographs of the individual graphene layers prior to, and after assembly into the heterostructure, and conventional dark-field electron micrographs of the samples assembled on TEM grids. Virtual apertures where then obtained by using threshold intensities to draw contours in the averaged diffraction patterns (see Supplemental Information section 1 for motivation), which were then used to obtain the masks associated with each overlap region. The intensity within these masked regions was then summed for each real space pixel to yield the virtual dark fields shown. In practice we used only the pixels close to the centers of each region (each region was down-sized by 25 percent) to minimize the bleed-in of intensity modulations originating from interference with other disks. Colored stacking order maps were then constructed from the virtual dark field images associated with the first and second order Bragg reflections using the provided bi-variate color-map, with red, blue, and green channels equal to the average intensity in the first order disks, second order disks, and the average of the red and blue channel values, respectively. Attribution of intensities to stacking order is rationalized in Supplemental Information sections 4--6. 

\subsection*{Twist Angle and Heterostrain Measurement} Twist angles were determined through triangulating the high-symmetry stacking orders in accordance with previous work. \cite{kazmierczak2021strain, van2022quantitative}. For AtA samples, the bright AAA stacking regions were identified by fitting the data to a series of Gaussians, the centers of which were triangulated using the Delaunay algorithm. The resultant moire wavelengths were then used to determine the twist angle and percent of heterostrain through fitting to the expressions provided in \cite{kerelsky2019maximized, van2022quantitative} using non-linear least squares. The twist angle associated with the larger moiré was calculated similarly from the bright tAA/AtA regions within the two-layer overlap virtual dark fields. For tAB samples, we instead inverted the virtual dark field and triangulated the centers of the ABC stacking regions. For samples containing a large portion of heterostrain and/or an insufficiently large field of view, the smallest viable moiré triangle was used to create a bound for the twist angle and percent heterostrain.

\subsection*{Analysis of Stacking Area Portions} 
Stacking area fractions were determined by thresholding the virtual dark field intensities into the categories illustrated in Supplemental Information Figures 6 and 9. This corresponds to region definitions of \ringonemacro $> 0.5$ and \ringtwomacro $< 0.5$ labeled AtB/tAB, \ringonemacro $> 0.5$ and \ringtwomacro $> 0.5$ labeled AtA/tAA, and those remaining SP. All statistical analyses and stacking order partitioning were performed after smoothing the virtual dark field images with a Gaussian filter of $\sigma = 0.5$ nm. For the AtA and tAB samples shown, this partitioning was applied directly to the virtual dark field image obtained from the three-disk overlap region. For the remaining samples, the geometric partitioning was first performed on the the two-layer overlap virtual dark field micrographs to obtain masks for the stacking order in the larger moire pattern. These masks were then applied to the three-layer virtual dark fields to isolate the regions of tAB and tAA stacking so that they could be independently analyzed as in the case of AtA and tAB samples.

\section*{Data Availability}
The data supporting the findings of this study are available within the Article and its Supplementary Information files. Datasets can be found at \href{https://doi.org/10.5281/zenodo.4459669}{https://doi.org/10.5281/zenodo.4459669}. 

\section*{Code Availability}
The code developed for data analysis in this study is available within the TrilayerTEM sub-directory at \href{https://github.com/bediakolab/bediakolab\_scripts}{https://github.com/bediakolab/bediakolab\_scripts}.

\printbibliography

\newpage

\begin{figure*}
    \centerline{\includegraphics[width=\textwidth]{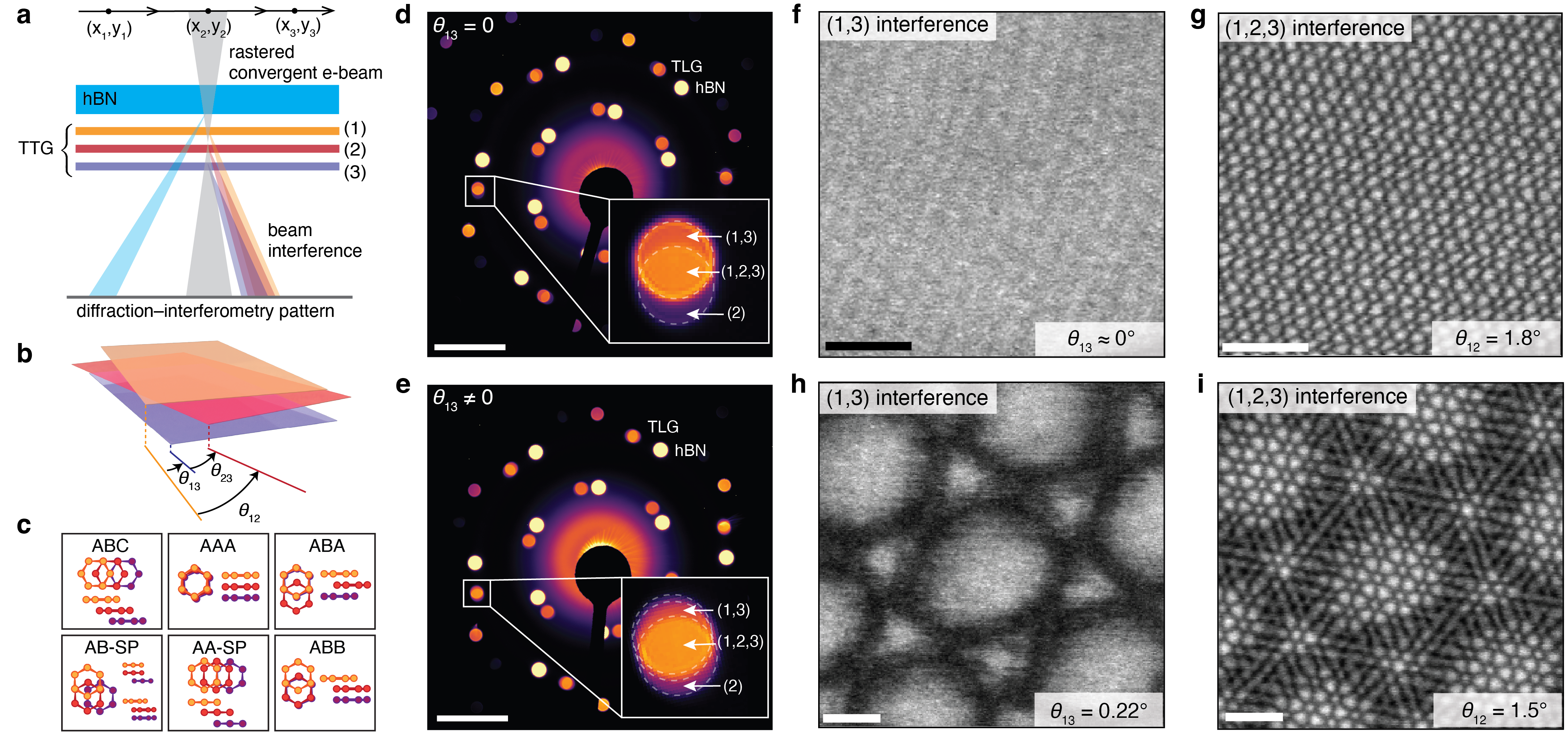}}
    \caption{ \textbf{Interferometric 4D-STEM dark field imaging of selected interfaces.} (\textbf{a}) Schematic of the 4D-STEM approach, wherein beam interference is used to extract stacking order. (\textbf{b}) Schematic illustrating the twist angle, $\theta$, and layer numbering conventions used to label the graphene trilayers. (\textbf{c}) Illustrations of various high-symmetry stacking configurations realized within twisted trilayer moirés. (\textbf{d,e}) Average convergent beam electron diffraction patterns for trilayers with $\theta_{13} \approx 0\degree$ \textbf{(d)} and $\theta_{13} = 0.22^{\degree}$ \textbf{(e)}. Overlapping TTLG Bragg disks are highlighted in the insets. Attribution of each Bragg disk to a layer is motivated in Supplemental Information section 3. \textbf{(f,h)} Virtual dark field images corresponding to the overlap of layers 1 \& 3. \textbf{(g,i)} Virtual dark field images corresponding to the overlap of all three layers. Scale bars are 1 nm$^{-1}$ and 25 nm for reciprocal (\textbf{d,e}) and real space (\textbf{f--i}), respectively. }
\end{figure*}

\begin{figure*}
    \centerline{\includegraphics[width=\textwidth]{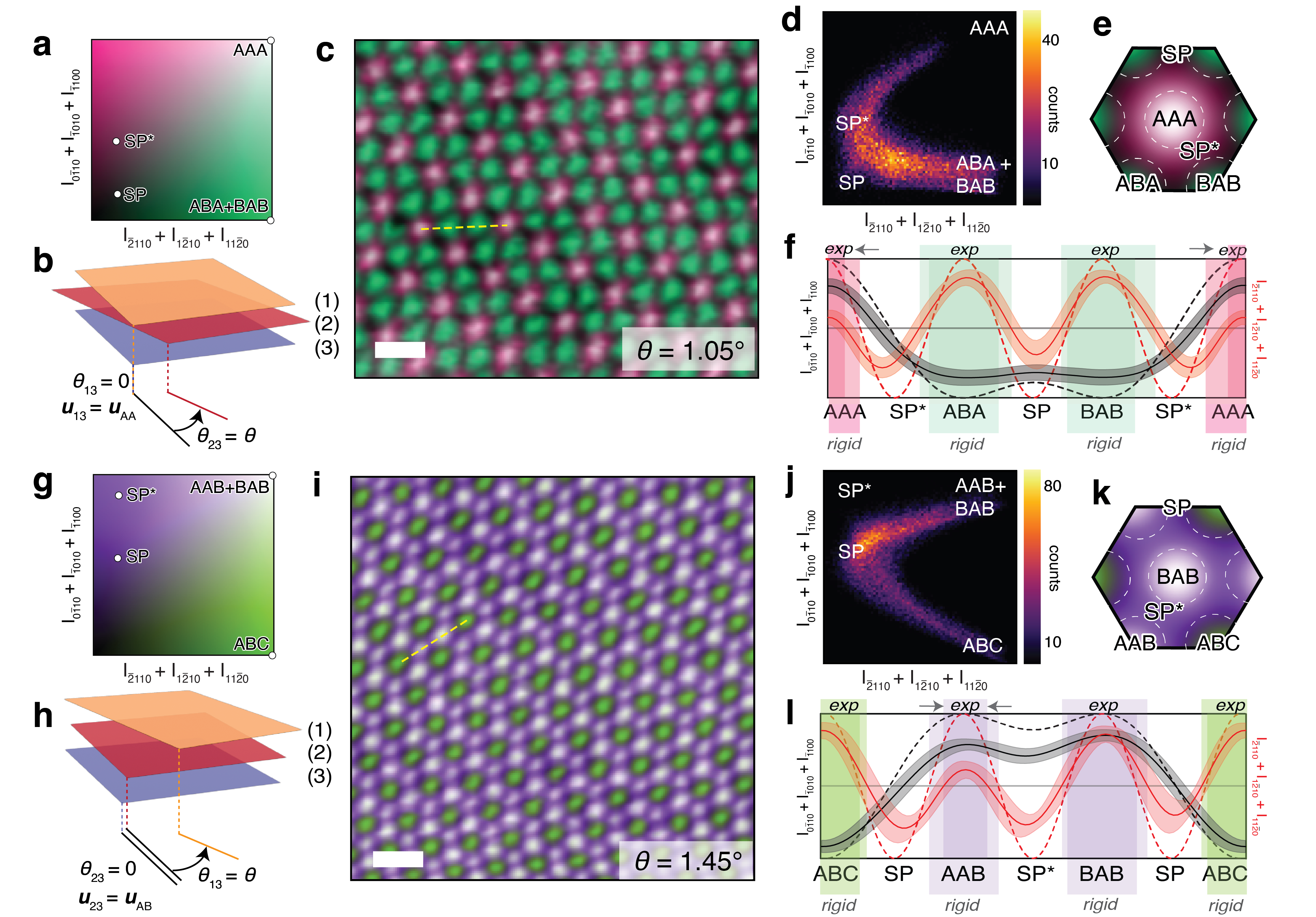}}
    \caption{ \textbf{Reconstruction in AtA and tAB trilayers. }(\textbf{b, h}) schematic illustrating the layer alignment in a AtA and a tAB trilayer. (\textbf{a, g}) Legends illustrating how color correlates with the average first and second order Bragg disks intensities. Overlaid points are the intensities of high symmetry stacking orders obtained via multislice, see Methods. (\textbf{c, i}) Maps of local stacking order for AtA and tAB trilayers, cropped to exclude biasing from sample drift. (\textbf{d, j}) Histograms illustrating the relative prevalence of each stacking configuration. Note that since the intensity does not depend linearly on the stacking order, a rigid bilayer will not display a uniform distribution of intensities (see Supplemental Information sections 4--5). (\textbf{e, k}) Schematics illustrating the anticipated variation in local stacking order expected for a rigid structure, obtained via the expressions provided in Supplemental Information section 5. The convention is used that AAA, ABA, SP*, and SP stacking denote interlayer offsets of $u_{12} = (0,0), (a_0/\sqrt{3},0), (a_0(2\sqrt{3})^{-1}, 0), $ and $(0, a_0/2)$ respectively in Cartesian coordinates where $a_0$ is the lattice constant. Similarly for the tAB samples, AAA, ABC, SP*, and SP stacking denote interlayer offsets of $u_{13} = (0,0), (a_0/\sqrt{3},0), (a_0(2\sqrt{3})^{-1}, 0), $ and $(0, a_0/2)$ respectively.  (\textbf{f, l}) Intensity line-cuts corresponding to the average of all possible line-cuts equivalent to the dotted lines shown in (\textbf{c, i}) are given as solid lines. Shaded region represents the standard deviation and arrows denote the statistically significant contractions of AAA and AAB domains. Intensity variation expected for a rigid structure are given as dotted lines. Domain sizes are calculated from the full width at half max of \ringtwomacro (red) as highlighted, where the value of \ringonemacro (black) is used to distinguish between different high symmetry stacking orders (see Supplemental Information Section 6 for validation of this approach).  All scale bars are 25 nm. }
\end{figure*}

\begin{figure*}
    \centerline{\includegraphics[width=400pt]{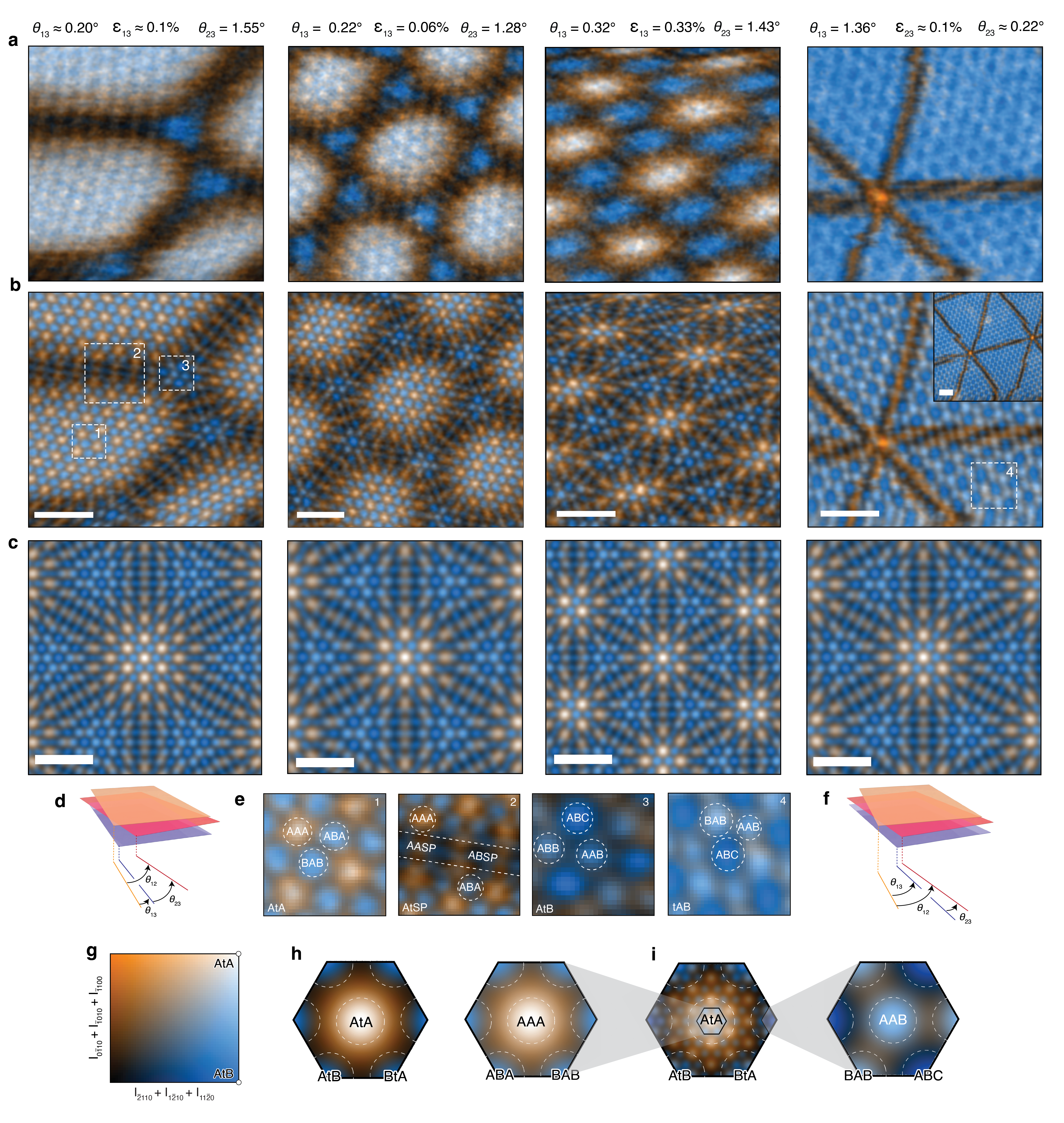}}
    \caption{ \textbf{Atomic Stacking in slightly misaligned TTLG.} (\textbf{a}) Maps of local atomic stacking from the larger moiré pattern only, corresponding to the local in-plane offset between between layers 1 and 3 in panels 1-3, and the local in-plane offset between layers  2 and 3  in panel 4. Colors shown correspond to the bi-variate colormap in (\textbf{g}), with accompanying expressions and simulations motivating the attribution of intensities to stacking orders as labeled here in Supplemental Information sections 4--6. (\textbf{b}) Local atomic stacking obtained from considering all three graphene layers. (\textbf{c}) Simulated stacking order maps for rigid moiré superlattice analogues of \textbf{b}, obtained from the expression given in Supplemental Information Section 5. All scale bars are 25 nm. (\textbf{d,f}) Schematics of layer alignment in TTLG with slightly misaligned layers. (\textbf{e}) Zoom-ins of the maps above illustrating the finer local modulation of stacking order within (\textbf{1}) AtA, (\textbf{2}) AtSP, (\textbf{3}) AtB, and (\textbf{4}) tAB regions. (\textbf{g}) Legend illustrating how color correlates with the average first and second order Bragg disks intensities for both the two and three layer interference patterns, with labeled locations of two layer high symmetry regions. \textbf{(h)} The expected intensity variation for an individual bilayer interface within the general trilayer structure obtained via the expression provided in Supplemental Information section 4. \textbf{(i)} Variation in first and second order Bragg disks intensities expected for a rigid twisted trilayer obtained via the expression provided in Supplemental Information section 5. Intensity relations are verified with multi-slice simulations in Supplemental Information section 6. }
\end{figure*}

\begin{figure*}
    \centerline{\includegraphics[width=\textwidth]{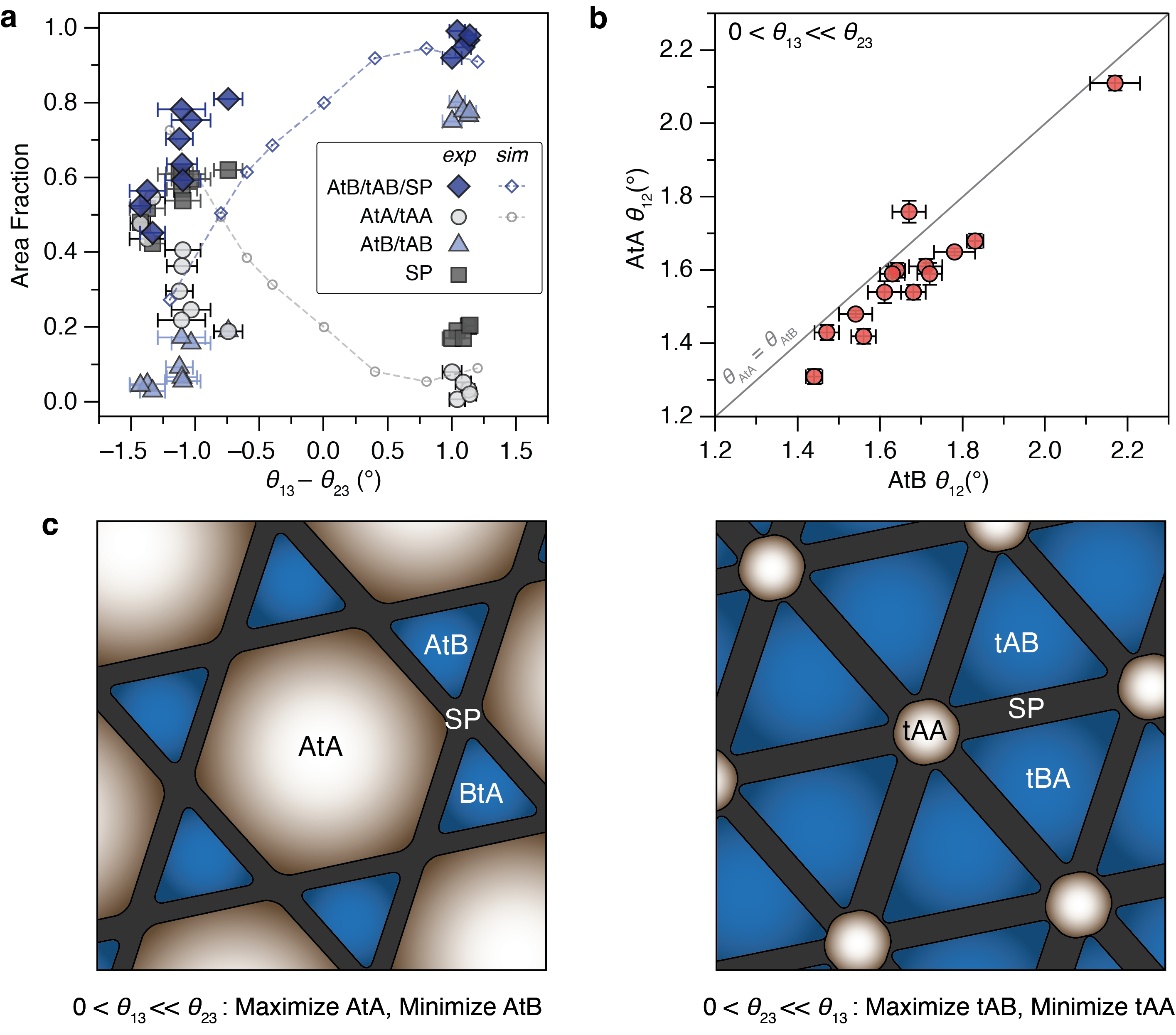}}
    \caption{ \textbf{Reconstruction patterns and trends in TTLG.} (\textbf{a}) Area fraction of atomic stacking domains from the larger moiré pattern only as a function of $\theta_{13}-\theta_{23}$. As both $\theta_{13}$ and $\theta_{23}$ are slightly variable for the samples discussed, additional plots of area fraction against $\theta_{13}$ and $\theta_{23}$ independently are provided in Supplemental Information Fig. 10. Experimental (\emph{exp}) data (corresponding to the maps shown in Fig. 3 and Supplemental Information Figs. 7-8) are compared with relaxation simulations (\emph{sim}). Area fractions associated with the experimental and simulated data were obtained following the procedure described in the methods (with regions of \ringtwomacro $> 0.5$ and \ringonemacro $< 0.5$ labeled AtB/tAB, \ringtwomacro $> 0.5$ and \ringonemacro $> 0.5$ labeled AtA/tAA, and those remaining SP) and by applying a threshold of 0.25 degrees to the local curl respectively (with further details in Supplemental Information section 9). We note that these methods result in functionally equivalent categorizations due to the small area and large intensity variation associated with the soliton regions where these thresholds partition the data. Error bar widths are the twist angle standard deviation from 53-112 data points for each point.  (\textbf{b}) Local twist angle associated with the smaller moiré within AtA and AtB domains. All points correspond to the regime where $\theta_{13} \ll \theta_{23}$. Twist angle determination is described in the methods. (\textbf{c}) Qualitative schematic illustrating the atomic reconstruction patterns (large moiré) observed for $\theta_{13} \ll \theta_{23}$ and $\theta_{23} \ll \theta_{13}$. Error bar widths are the twist angle standard deviation from 3-33 data points for each point. }
\end{figure*}

\begin{figure*}
    \centerline{\includegraphics[width=\textwidth]{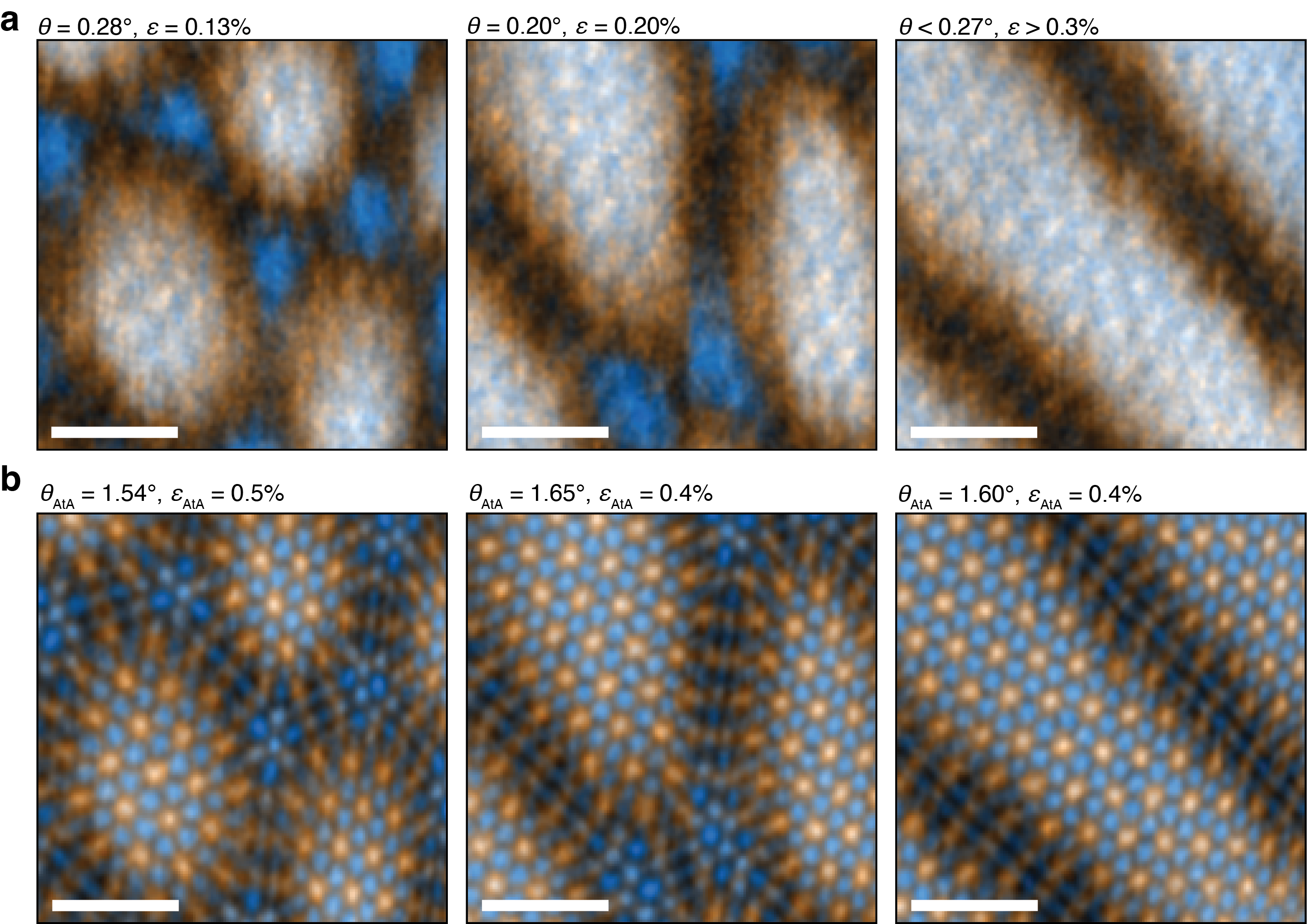}}
    \caption{ \textbf{Heterostrain Effects} (\textbf{a}) Maps of the modulation in local stacking order between layers 1 and 3 only for samples with an increasingly large percent of extrinsic heterostrain. (\textbf{b}) Corresponding maps of the local stacking order modulation obtained when considering all three graphene layers. Twist angles and percent heterostrains values and bounds were determined from fitting the size and asymmetry of the moiré triangles (see methods). All scale bars are 25 nm.}
\end{figure*}

\end{document}